\documentclass{an}
\usepackage{graphicx}
\usepackage{times}
\usepackage{fancyhdr}
\begin{document}

\title{The low-mass companion of GQ Lup}

\author{
E.W. Guenther $^1$, R. Neuh\"auser $^2$, G. Wuchterl $^2$, M. Mugrauer
$^2$, \\
A. Bedalov $^2$, P.H. Hauschildt $^3$}

\institute{$^1$ Th\"uringer Landessternwarte Tautenburg, 07778
Tautenburg, Germany \\ 
$^2$ Astrophysikalisches Institut und Universit\"ats-Sternwarte
Schillerg\"a\ss chen 2-3, 07745 Jena, Germany\\
$^3$ Hamburger Sternwarte, Gojenbergsweg 112, 21029 Hamburg Germany \\
}

\date{Received 2.8.2005; accepted 21.10.2005; published online}

\abstract{Using NACO on the VLT in the imaging mode we have detected an
object at a distance of only 0.7 arcsec from GQ Lup. The object turns
out to be co-moving. We have taken two K-band spectra with a resolution
of $\lambda / \Delta \lambda=700$.  In here, we analyze the spectra in
detail. We show that the shape of spectrum is not spoiled by differences
in the Strehl ratio in the blue and in the red part, as well as
differential refraction. We reanalyze the spectra and derive the
spectral type of the companion using classical methods. We find that the
object has a spectral type between M9V and L4V, which corresponds to a
$T_{eff}$ between 1600 and 2500 K. Using GAIA-dusty models, we find that
the spectral type derivation is robust against different
log(g)-values. The $T_{eff}$ derived from the models is again in the
range between 1800 and 2400 K.  While the models reproduce nicely the
general shape of the spectrum, the $^{12}CO$-lines in the spectrum have
about half the depth as those in the model. We speculate that this
difference might be caused by veiling, like in other objects of similar
age, and spectral class. We also find that the absolute brightness of
the companion matches that of other low-mass free-floating objects of
similar age and spectral type.  A comparison with the objects in USco
observed by Mohanty et al. (\cite{mohanty04b}) shows that the companion
of GQ Lup has a lower mass than any of these, as it is of later spectral
type, and younger. The same is as true, for the companion of AB Pic.  To
have a first estimate of the mass of the object we compare the derived
$T_{eff}$ and luminosity with those calculated from evolutionary
tracks. We also point out that future instruments, like NAHUAL, will
finally allow us to derive the masses of such objects more precisely.
}

\correspondence{guenther@tls-tautenburg.de}

\maketitle

\section{Introduction}

Now more than 160 extrasolar planets have been discovered indirectly by
means of precise the radial velocity measurements of the host stars. At
least for the 6 transiting planets the planetary nature of these objects
is confirmed (e.g. Charbonneau et al. \cite{charbonneau00}). In two
additional cases the planetary nature of the orbiting objects is
confirmed astrometrically (Benedict et al. \cite{benedict02}).  In many
other cases, astrometric measurements are at least precise enough to
rule out binary star viewed almost face on.  

A statistical analysis shows that the observed frequency of solar-like
stars having planets with a minimum mass $\geq$ $0.3\,M_{\rm Jupiter}$
orbiting at distances of $\leq 5$ AU is 9\% (Lineweaver \& Grether
\cite{lineweaver03}).  It is thus quite surprising that brown dwarfs are
very rare as close companions to normal stars, in contrast to planets
and stellar companions.  The lack of brown dwarfs as companions is thus
often referred as the brown dwarf desert.  Marcy et al. (\cite{marcy03})
estimate from their radial velocity (RV) survey of old, solar-like stars
that the frequency of brown dwarfs with 3 AU of the host stars is only
$0.5\pm0.2\%$, and thus much smaller than the frequency of planets, or
the frequency of binaries. This result is recently confirmed by a radial
velocity survey of stars in the Hyades which, combined with AO-imaging
also shows that the number of companions with masses between 10 $M_{\rm
Jupiter}$ and 55 $M_{\rm Jupiter}$ at distances $\leq 8$ AU is $\leq$
2\% (Guenther et al. \cite{guenther05}).  Studies by Zucker \& Mazeh
(\cite{zucker01}) show that the frequency of close companions drops off
for masses higher than 10 $M_{\rm Jupiter}$, although they suspect there
is still a higher mass tail that extends up to probably 20 $M_{\rm
Jupiter}$. From the currently known ``planets'' 15 have an $\rm
m\,sin\,i$ between 7 and 18 $\,M_{Jupiter}$. It has been argued by Rice
et al. (\cite{rice03}) that these massive planets do not form by core
accretion, because the host stars do not show enhanced metallicity,
unlike stars hosting planets of lower mass.  Wide companions (e.g. $d
\geq 50$ AU) are detected by means of direct imaging. Unfortunately this
means that their masses can only be estimated by comparing their
temperature and luminosities with evolutionary tracks.  In the case of
these pairs, the situation is possibly different, as direct imaging
campaigns probably have turned up 11 brown dwarfs orbiting normal
stars. The result of all search programs for objects in TWA-Hydra,
Tucanae, Horologium and the $\beta$ Pic region is that the frequency of
brown dwarfs at distances larger than 50 AU is $6\pm4\%$ (Neuh\"auser et
al. \cite{neuhaeuser03}).  This result implies that the frequency of
wide binaries consisting of a brown dwarf and a star is much higher than
that of close binaries, or it means that there are serious problems with
the tracks.

Recently, three very low-mass companions have been identified that could
possibly even have masses below 13 $M_{\rm Jupiter}$. 2MASSWJ
1207334-393254 is a brown dwarf with a spectral type M8V.  A co-moving
companion has been found which is located at a projected distance of 70
AU (Chauvin et al. \cite{chauvin05a}).  The companion has a spectral
type between L6 and L9.5.  Assuming that 2MASSWJ 1207334-393254 is a
member of the TW Hydra association, and assuming an age $8^{+4}_{-3}$
Myr, the mass of the primary is 25 $M_{Jupiter}$.  Using the non-gray
models from Burrows et al. (1997), the authors estimate the mass of the
companion as 3 to 10 $M_{\rm Jupiter}$. $AB\,Pic$ also has a very
low-mass companion (Chauvin et al. \cite{chauvin05b}).  $AB\,Pic$ is a
K2V star in the Tucana-Horologium association.  The age is estimated as
$\sim$ 30 Myr. The co-moving companion with a spectral type of L0 to L3
is located at the projected distance of 260 AU from the primary. The
K-band spectrum of the companion shows the NaI doublet at 2.205 and 2.209
$\mu m$. For this object, the authors give a mass estimate between 13
and 14 $M_{\rm Jupiter}$. The third such object is GQ Lupi which will be
discussed here.

\section{GQ Lup}

\begin{figure}
\resizebox{\hsize}{!}
{\includegraphics[width=0.55\textwidth, angle=-90]{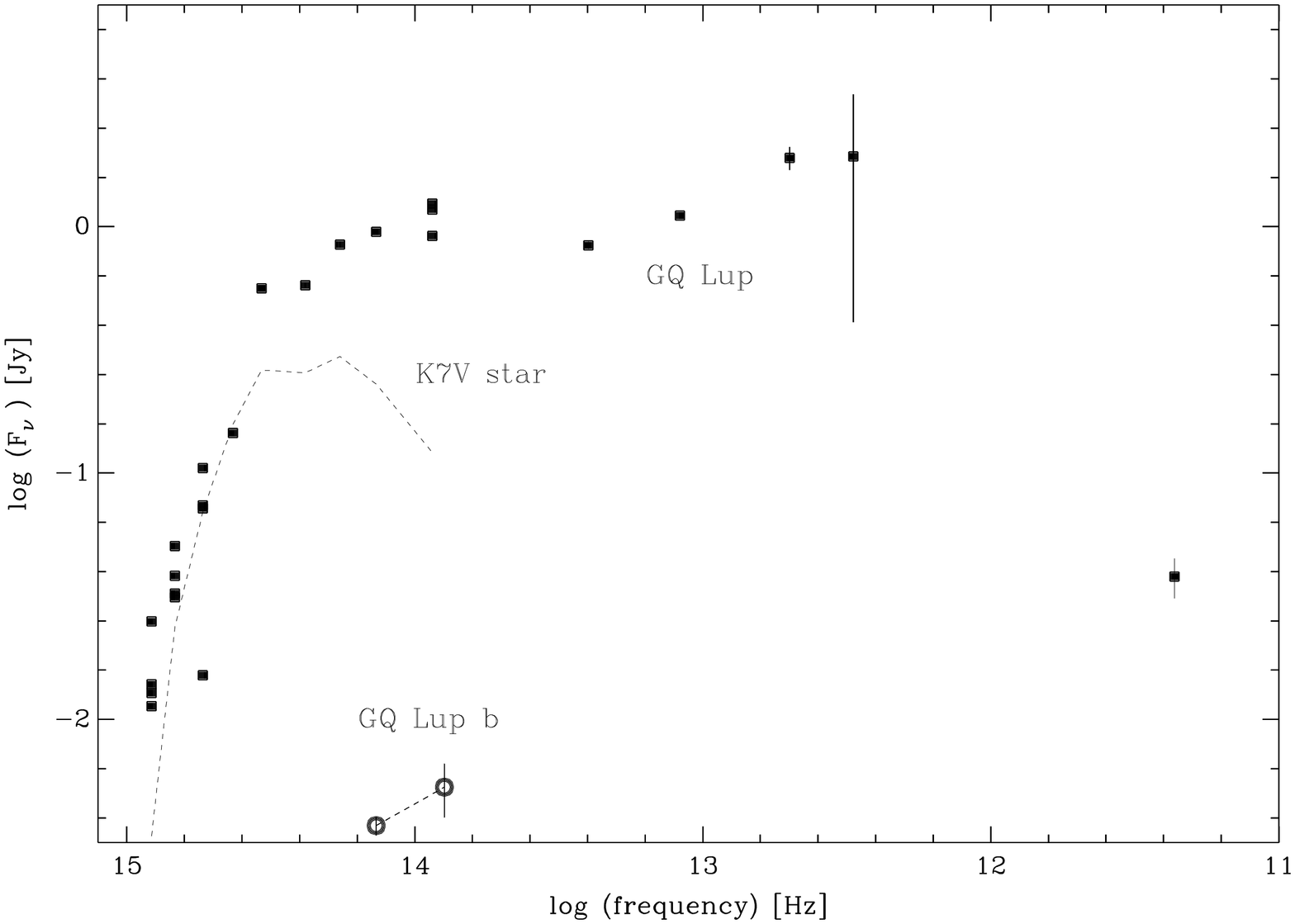}}
\caption{The figure show the spectral energy distribution of GQ Lup 
as derived by using all photometric measurements taken from the literature.
Also shown are the two photometric measurements of GQ Lup b, and a
K7V star with a diameter of 1.5 $R_\odot$ located at a distance of 140 pc.}
\label{GQ_Lup_SED}
\end{figure}

GQ Lup is a classical T Tauri star of YY Orionis type located in the
Lupus I star-forming region. Quite a number of authors have determined
the distance to this star-forming region: Hughes et
al. (\cite{hughes93}) find $140\pm20$~pc, Knude \& H\o g
(\cite{knude98}) 100~pc, Nakajima et al. (\cite{nakajima00}) 150 pc,
Satori et al. (\cite{satori03}) 147, Franco et al. (\cite{franco02}) 150
pc, de Zeeuw et al. (\cite{zeeuw99}) $142\pm2$~pc, and Teixeira et
al. (\cite{teixeira00}) 85~pc but note that 14 stars of this group have
measured parallax-distances, which are are on average 138~pc. The most
likely value for the distance thus is 140~pc, which will be used in the
following. The spectral type of GQ Lup is K7V. Batalha et
al. (\cite{batalha01}) find a veiling between 0.5 and 4.5 and an
extinction $A_V$ of $0.4\pm0.2$ mag, which implies an $A_K=$
$0.04\pm0.02$ mag, and $A_L=$ $0.02\pm0.01$ mag.  Using spectra taken
with HARPS, we derive a $v\,sin\,i$ of $6.8\pm0.4$ $km\,s^{-1}$,
assuming a Gaussian turbulence velocity of 2 $km\,s^{-1}$, and assuming
a solar-like center to limb variation.  The broad-band energy
distribution of GQ Lup is shown in Fig.\,\ref{GQ_Lup_SED} together with
a K7V star of 1.5 $R_\odot$ located at 140 pc. In the optical, the data
fits nicely to a star with low to medium veiling, as observed. In the
infrared, a huge excess due to the disk is seen.

% original version
% \begin{figure}
% \resizebox{\hsize}{!}
% {\includegraphics[width=0.45\textwidth, angle=-90]{EGuenther_fig1.ps}}
% \caption{The figure show the spectral energy distribution of GQ Lup 
% as derived by using all photometric measurements taken from the literature.
% Also shown are the two photometric measurements of GQ Lup b, and a
% K7V star with a diameter of 1.5 $R_\odot$ located at a distance of 140 pc.}
% \label{GQ_Lup_SED}
% \end{figure}

\section{The spectrum of the companion}

We detected a faint companion at a distance of $732.5\pm3.4$~mas with a
positional angle of $275.45\pm0.30^o$ (Neuh\"auser et
al. \cite{neuhaeuser05}). As described in more detail in Mugrauer \&
Neuh\"auser (\cite{mugrauer05}), using our own imaging data, as well as
data retrieved from the HST and SUBARU archive, it was shown that the
pair has common proper motion at significants-level of larger than 7
$\sigma$.  

After this question is solved, the next question to solve is, what the
companion is.  Using NACO, we obtained two spectra of the companion. The
first spectrum was taken on August 25, 2004, the second on September 13,
2005. The first spectrum had a S/N-ratio of only 25, that is why it was
repeated. The second spectrum has a S/N-ratio 45. For our observations
we used $S54\,SK$-grism and a slit width of 172~mas which gives a
resolution of about $\lambda / \Delta \lambda=700$. Because the Strehl
ratio, as well as the refraction depends on wavelength, the flux-loss in
the blue and in the red part of the spectrum may differ if a very narrow
slit is used. However, since we used a relatively wide slit, and
observed airmass 1.24, and 1.30 respectively, this effect is only
1.5\% for the wavelength region between 1.8 and 2.6 $\mu m$.

\begin{figure}
\resizebox{\hsize}{!}
{\includegraphics[width=0.45\textwidth, angle=0]{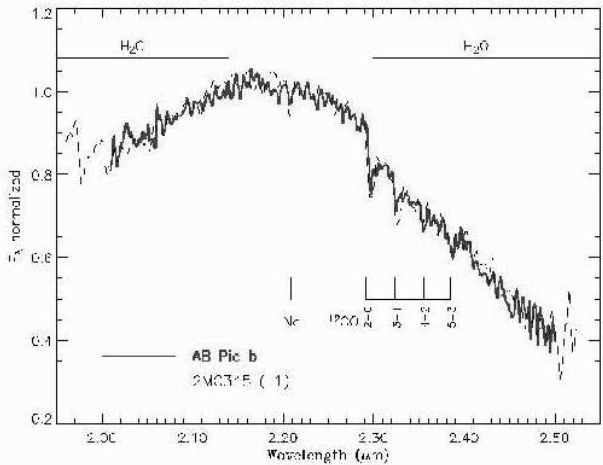}}
\caption{Spectrum of the companion of AB Pic taken from Chauvin et
al. (\cite{chauvin05b}). These authors assign a spectral type L0 to L3
and an age of 30 Myr to this companion.}
\label{AB_Pic_b_spectrum}
\end{figure}

There are several classical methods as to derive the spectral types of
late-type objects from spectra taken in the K-band.  Using the K1-index
from Reid et al. (\cite{reid01})
(K1=[2.10-2.18]-[1.96-2.04]/(0.5*[2.10-2.18]+[1.96-2.04]); Sp -2.8 +
K1*21.8), we find spectral types in the interval M9V to L3V, using the
two spectra and using different methods for the flux calibration.  Using
the $H_20-D$-coefficient from McLean et al. (\cite{mclean03}) which is
simply the flux ratio between $1.964$ to $2.075$ $\mu m$, we derive
spectral types in the range between L2V to L4V. However, this
coefficient is known to have an accuracy of only one spectral class. In
order to be on the save side, we thus estimate the spectral type to be
between M9V to L4V. Another piece of evidence is the NaI lines at 2.2056
and 2.2084 $\mu m$. These line vanishes at a spectral types later than
L0V. Unfortunately, there is a telluric band between 2.198 and 2.200
$\mu m$, which is difficult to distinguish from the NaI lines in a low
resolution spectrum. We thus can only give an upper limit of 3 \AA , for
the equivalent width of the NaI doublet. Using the conversion from
spectral type to $T_{eff}$ from Basri et al. (\cite{basri00}),
Kirkpatrick et al.  (\cite{kirkpatrick99}), and Kirkpatrick et al.
(\cite{kirkpatrick00}), this range of spectral types corresponds to
$T_{eff}$-values in the range between 1600 to 2500 K.

\begin{figure}
\resizebox{\hsize}{!}
{\includegraphics[width=0.45\textwidth, angle=-90]{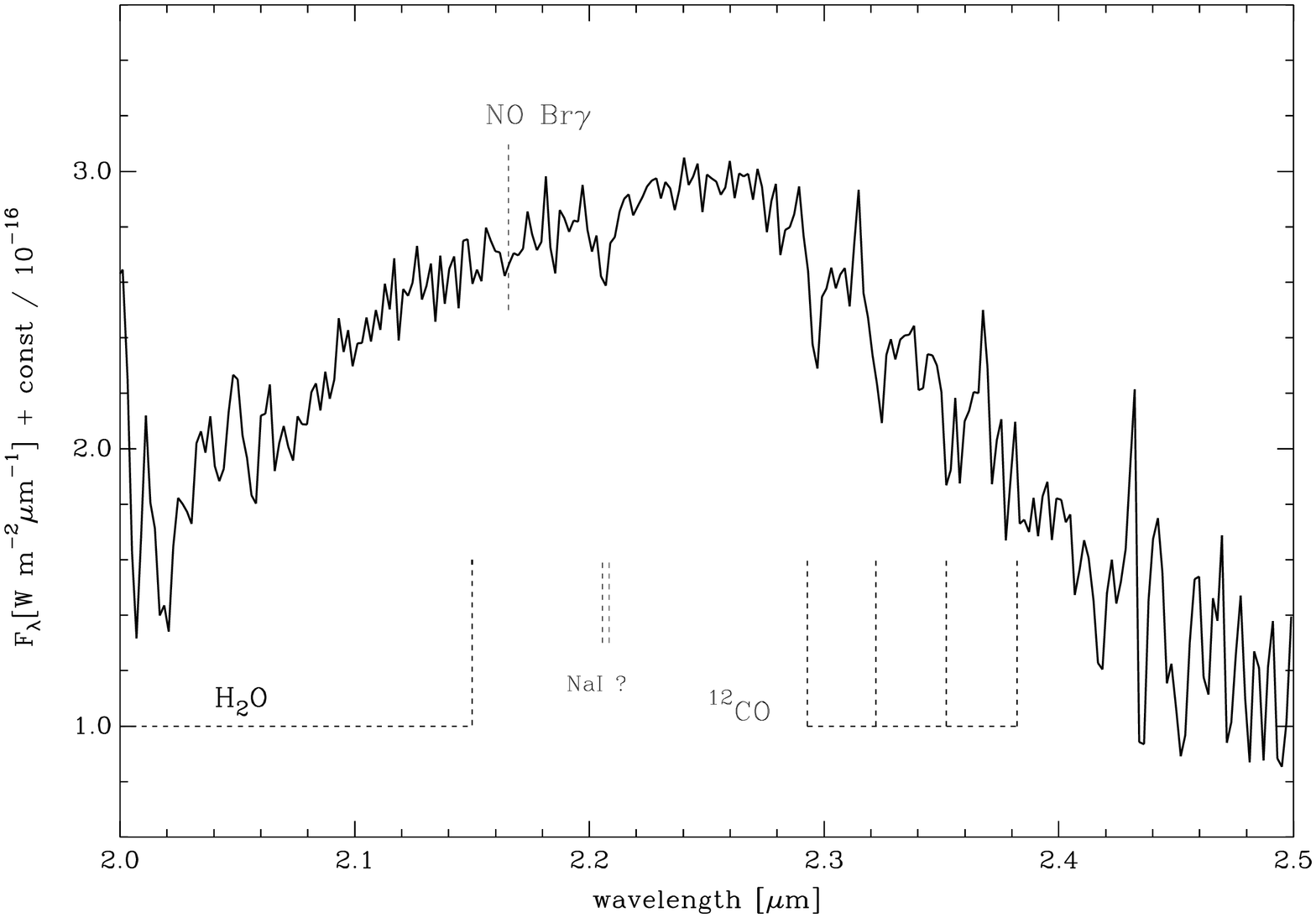}}
\caption{Spectrum of the companion of GQ Lup. The figure has exactly
the same scale as Fig.\,\ref{AB_Pic_b_spectrum}. As can easily be seen, the
spectra of the two objects are quite similar. The depth of the CO-lines
is also similar.
We assign a spectral type M9 to L4.
}
\label{gq_lup_comp_flux}
\end{figure}

The expected K-L'-colours of an object with a spectral type M9V to L4V
are between 0.5 and 1.2 mag, which matches reasonably well the derived
K-L'-colour of $1.4\pm0.3$ of the companion (Golimowski et
al. \cite{golimowski04}). Using the extinction to the primary, and
assuming a distance of 140 pc, we derive from the observed brightness of
$m_Ks=13.1\pm0.1$, and $m_{L'}=11.7\pm0.3$, absolute magnitudes of
$M_Ks=7.4\pm0.1$ and $M_{L'}=6.0\pm0.3$ mag for the companion
(Fig.\,\ref{GQ_Lup_SED}).  Old M9V to L4V objects have $M_K$-values
between 9.5 to 12 mag and $M_{L'}$-values between 9.8 and 10.5 mag. The
companion thus is much brighter than old M, or L-dwarfs (Golimowski et
al. \cite{golimowski04}). When discussing the brightness of the
companion, we have to keep in mind that there are three additional
effects that may lead to large absolute magnitudes, apart from the young
age of the object: The first one simply is that it could be a binary.
The second is that the distance could be much smaller than 140~pc. The
third possibility is that the brightness is enhanced due to accretion
and a disk, like in T Tauri stars. In this respect it is interesting to
note that objects of similar age and spectral type often have disks and
show signs of accretion. Typical accretion rates are about $10^{-11}$
$M_\odot yr^{-1}$ (Liu, Najita, Tokunaga \cite{liu03}; Natta et
al. \cite{natta04}; Mohanty et al. \cite{mohanty04a}; Mohanty et
al. \cite{mohanty05a}; Muzerolle et al. \cite{muzerolle05}). Clear signs
of accretion are observed even down to the planetary-mass regime at
young ages (Barrado y Navascu\'es \cite{barrado02}). The fact that we do
not see the $Br_\gamma$-line in emission does not speak against the
accretion hypothesis, as the flux of this line is correlated with the
accretion rate, and at $10^{-11}$ $M_\odot yr^{-1}$, we do not expect to
see it (Natta et al. \cite{natta04}). The accretion hypothesis is
further supported by the fact that objects with spectral types of late M
in Taurus have $K_s-L'$-colours up to 1.2 mag, and absolute luminosities
of $M_{K}=6$ to 7, and $M_{L'}\sim 6.0$. The large luminosities and red
colours of these objects are usually interpreted as being caused by
disks and accretion (Liu, Najita, Tokunaga \cite{liu03}; Luhmann
\cite{luhman04}). The absolute magnitudes of the companion of AB\,Pic of
$M_J=12.8^{+1.0} _{-0.7}$, $M_H=11.3^{+1.0} _{-0.7}$, $M_K=10.8^{+0.9}
_{-0.7}$ are also quite similar to the of the companion of GQ\,Lup.
Thus, the companion of GQ Lup is quite a normal for an object of its
age, and we should keep in mind that it is likely that there is a disk,
and accretion.

\section{Comparing the spectrum with GAIA-dusty models}

Up to now we have compared the spectrum of the companion of GQ Lup with
spectra of old brown dwarfs which have a $log(g)\sim\,5.0$. Thus, one
may wonder, whether this causes a problem for the determination of the
spectral type.  In order to derive $T_{eff}$ it would be better to
compare the observed spectrum with spectra of different log(g). The only
way to do this, is to compare the observed spectrum with model
calculations. To do this, we use the GAIA-dusty models.

\begin{figure}
\resizebox{\hsize}{!}  {\includegraphics[width=0.45\textwidth,
angle=-90]{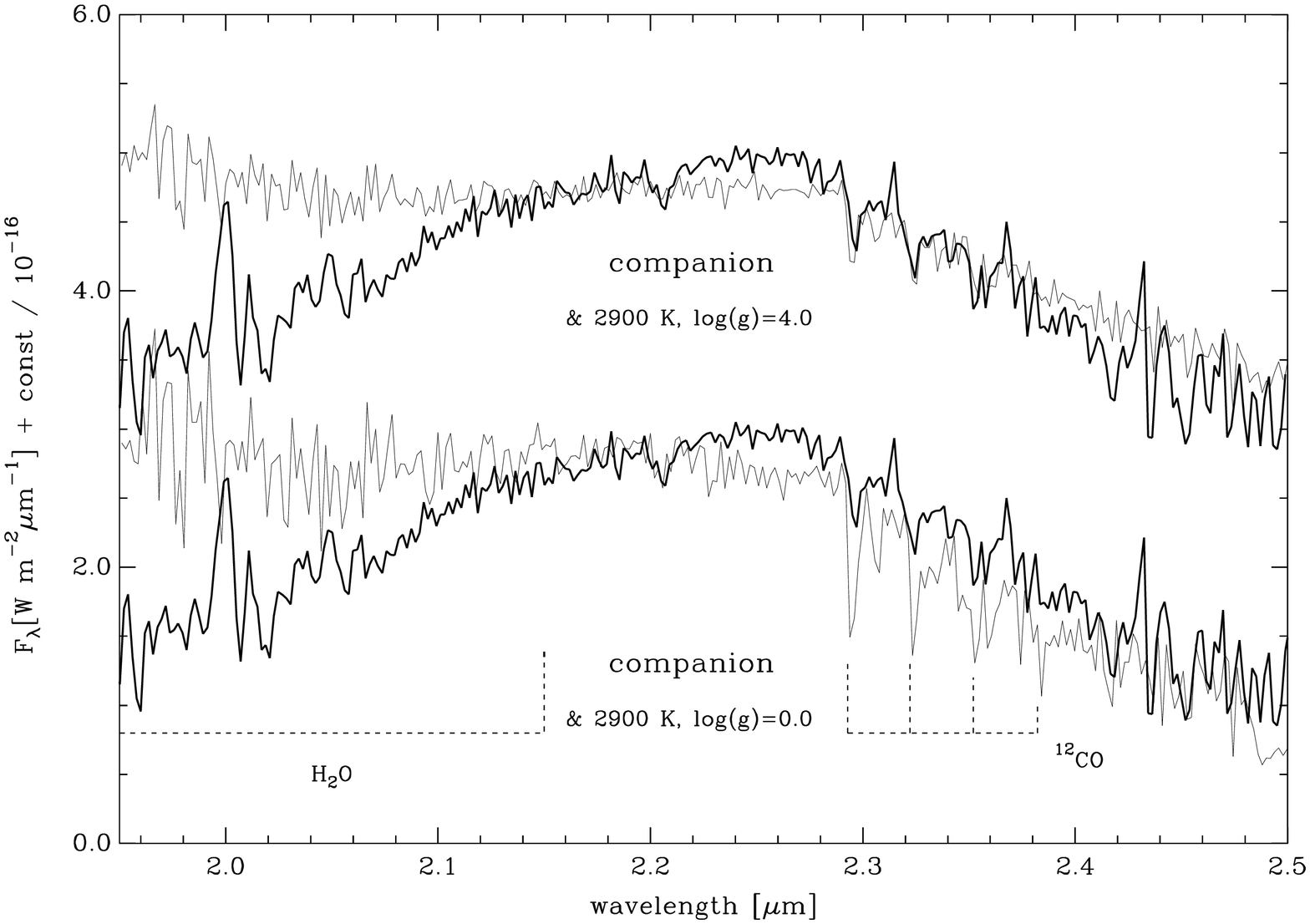}}
\caption{Flux calibrated spectrum of the companion of GQ
Lup. The thick line is the observed spectrum, the thin lines are models
calculated for $T_{eff}=$ 2900~K and log(g)=0, and log(g)=4.0. Clearly,
theses model do not fit to the data.  The object must be cooler than
that.}
\label{gq_lup_comp_flux_2900}
\end{figure}
 
Fig.\,\ref{gq_lup_comp_flux_2900} shows the flux-calibrated spectrum
together with two models. Both are calculated for a temperate of 2900~K.
One is for log(g)=0 and the other for log(g)=4.0. While the model with
log(g)=4.0 reproduces nicely the $^{12}CO$lines and to the NaI doublet
at 2.205 and 2.209 $\mu m$, it does fit to the $H_2O$-band in the
spectrum. Clearly, the object must be cooler than this. Also, if the
$T_{eff}$ were 2900~K, the radius of the object would be $\sim 1.0$
$R_{Jupiter}$. Which does not seems plausible for a very young low-mass
object.

\begin{figure}
\resizebox{\hsize}{!}
{\includegraphics[width=0.45\textwidth,
angle=-90]{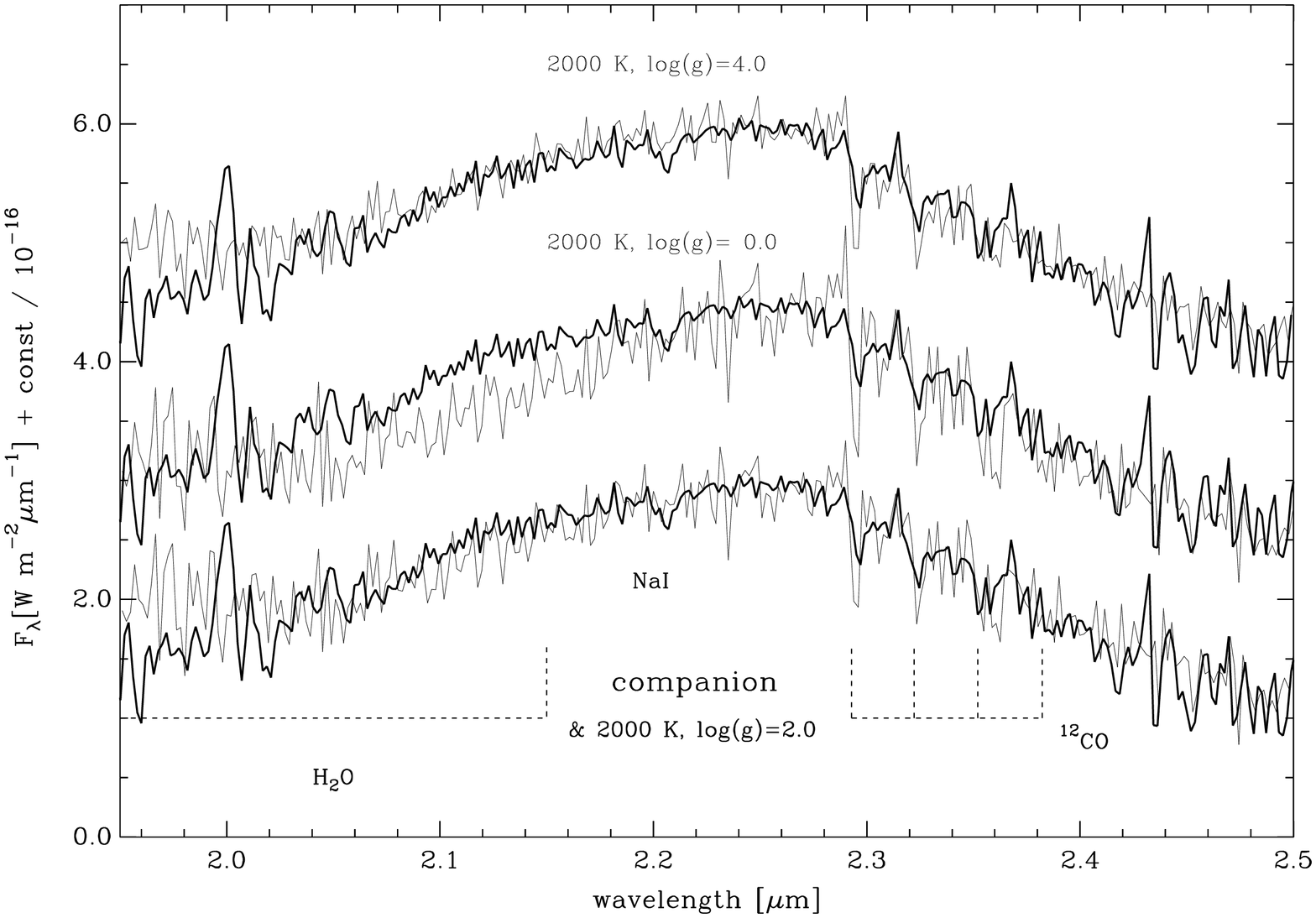}}
\caption{Flux calibrated spectrum Spectrum of the companion of GQ Lup. The 
thick line is the observed spectrum, the thin lines are models calculated for
$T_{eff}=$ 2000~K and log(g)=0, log(g)=2.0, log(g)=4.0. Clearly, theses model 
fit much better than the ones in Fig.\,\ref{gq_lup_comp_flux_2900}.}
\label{gq_lup_comp_flux_2000}
\end{figure}

Fig.\,\ref{gq_lup_comp_flux_2000} shows the flux-calibrated spectrum
together with two models calculated for a $T_{eff}$ of 2000~K. Judging
just from the shape of the spectrum, the three models almost perfectly
match the observed spectrum. The fit seems to be better for the two
models with log(g)=2.0 and log(g)=4.0. We can do this comparison a
little more quantitatively.  However, given the cross-talk between
log(g) and $T_{eff}$, and given that only a spectrum with a resolution
of $\lambda / \Delta \lambda$ of 700 is available, the currently
achievable accuracy of the determination of log(g) and $T_{eff}$ is
rather limited.  We find that $T_{eff}$-values in the range between 1800
and 2400~K, and log(g)-values between 1.7 bis 3.4 give good fits, in
excellent agreement with the previous temperature estimate. However, as
can easily bee seen, the $^{12}CO$-lines are always a factor two deeper
in the model than in the spectrum. If we assume that this difference is
caused by veiling due to the presence of the disk, the radius of the
companion would be 1.2 to 1.3 $R_{Jupiter}$. If we assume that there is
no veiling, the object would have a radius of 1.7 to 1.8
$R_{Jupiter}$. It is interesting to note that the depth of the
$^{12}CO$-lines in the spectrum of GQ\,Lup is the same as in the case of
the companion of AB Pic. This means that either both have the same
veiling, or the $^{12}CO$-lines in the models are too deep
(Fig.\,\ref{AB_Pic_b_spectrum}).

\section{Putting the object into perspective}

\begin{figure}
\resizebox{\hsize}{!}  {\includegraphics[width=0.45\textwidth,
angle=-90]{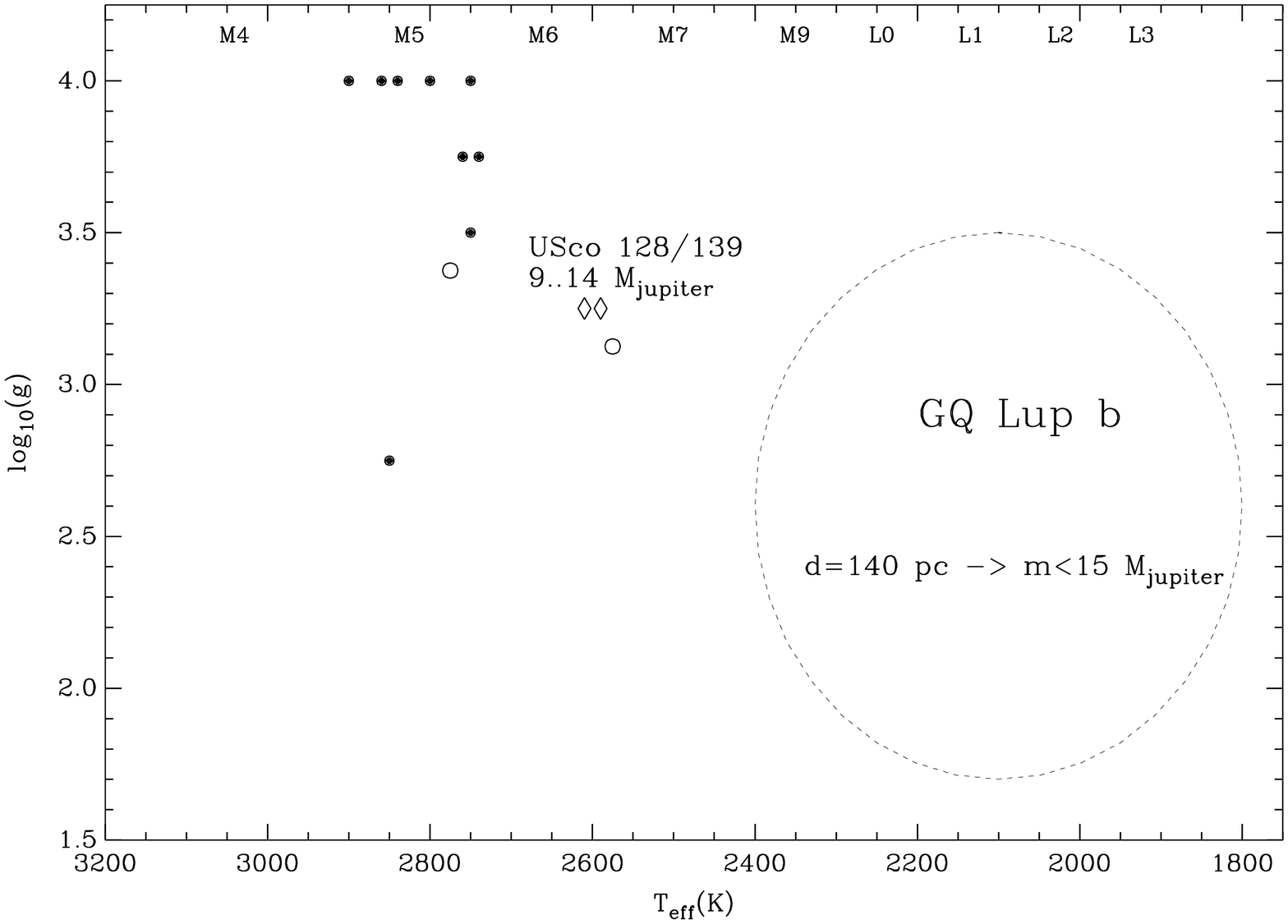}}
\caption{Comparing the $T_{eff}$ and the log(g)-values obtained for the
companion of GQ Lup with the values obtained for objects in USco by
Mohanty et al. (\cite{mohanty04b}) and Mohanty, Jayawardhana, Basri
(\cite{mohanty04c}).  The mass of the companion of GQ Lup must have a
mass lower than that of USco 128 and USco 130, because it has a later
spectral type and is younger. In these papers the authors give masses of
9 to 14 $M_{Jupiter}$ for USco 128 and USco 130, however as mentioned at
this conference, new values of oscillation strength of the TiO-lines
imply higher masses.}
\label{logg_Teff}
\end{figure}

The problem in giving a mass for the companion is that there is not a
single object with an age of about one Myr and such a late spectral type
where the mass has been determined directly.  Mohanty et
al. (\cite{mohanty04b}) attempted to do this by deriving the log(g) and
$T_{eff}$-values for late type objects on USco. These objects have an
age of about 5 Myr. For the analysis they used spectra with $\Delta
\lambda / \lambda=31\,000$ in the wavelength-range between 6400 and 8600
\AA . For USco 128 and USco 130, which have a spectral type of M7 and
M7.5, they find log(g)-values of 3.25 (Mohanty et al. \cite{mohanty04b};
Mohanty, Jayawardhana, Basri \cite{mohanty04c}).  With these values,
they find masses for these objects of 9 to 14 $M_{Jupiter}$. However,
during this meeting it was mentioned by the authors that the
log(g)-values are possibly too small by 0.5 dex (Mohanty
\cite{mohanty05b}). This would increase the masses of these objects to
$\geq 20$ $M_{Jupiter}$. In any case, the mass of the companion of GQ
Lup must be lower than that of USco 128 and USco 130, as it has a later
spectral type and is younger than these (Fig.\,\ref{logg_Teff}). Because
the companion of GQ Lup has the same spectral type as the companion of
AB Pic but is younger, it must have lower mass than it.

Given the cross-talk between log(g) and $T_{eff}$, and given that
we have a spectrum with a resolution of only $\lambda / \Delta
\lambda=700$, it is currently not possible to constrain the log(g)
sufficiently well to give a mass. For a radius of 1.2 to 1.3
$R_{Jupiter}$, a log(g) of $\leq3.7$ would imply a mass $\leq 13$
$M_{Jupiter}$. Similarly, if we assume that there is no veiling, a
log(g) of $\leq 3.4$ would imply a planetarial mass.

The problem when using evolutionary tracks for objects at very young
ages is that the brightness and temperature of the objects depend on the
history of the accretion. This means that in principle, the evolutionary
tracks from Burrows et al. (\cite{burrows97}) and Baraffe et
al. (\cite{baraffe02}) should not be used at such a young age. However,
it is still worth-while to have a look at these in order to have an
idea. Although an isochrone for $10^6$ years is not even shown in
Burrows et al. (\cite{burrows97}), the $3\,10^6$ year-isochrone leads to
a mass of 3 to 9 $M_{Jupiter}$, with a $T_{eff}$ of 1800 and 2400~K.
Similarly, we read of a mass between 3 to 16 $M_{Jupiter}$ from the
Fig.~2 in Baraffe et al. (\cite{baraffe02}).

% Mbol=10.57 oder 10.78
% Mittel 10.68 pm 0.145
% log(L7Lodot)-2.3761 +/-0.055

Burrows et al. (\cite{burrows97}) and Baraffe et al. (\cite{baraffe02})
also give the luminosity for objects at different ages. We may also try
to use this result for estimating the masses. According to Golimowski et
al. (\cite{golimowski04}) the bolometric correction $BC_K$ is
$3.17\pm0.06$ and $3.38\pm0.06$ for objects with spectral types of M9
and L4V, respectively. With $M_K=7.4\pm0.1$, this gives an
$M_{bol}=10.7\pm0.2$, or $log(L/L_\odot)=-2.38\pm0.08$, assuming a
distance of 140~pc (that is, not taking the error of the distance into
account), and assuming that there is no contribution from the disk, or
accretion. If we assume such a contribution, the luminosity goes down to
$log(L/L_\odot)=-2.7$. If we further assume that the distance would be
only 100 instead of the canonical 140~pc, we would obtain only
$log(L/L_\odot)=-3.0$. For these three assumptions, we derive masses of
about 20, 15 and 7 $M_{Jupiter}$ using Burrows et
al. (\cite{burrows97}), for the three hypothesis respectively.  Using
Baraffe et al. (\cite{baraffe02}) we find values of about 30, about 15,
and 10 $M_{Jupiter}$, or so.
 
Now in progress are models which take the formation of the objects into
account. Hubickyj, Bodenheimer, and Lissauer (\cite{hubickyj04}) model
the formation of giant planets via the accretion of planetesimals and
subsequent capture of an envelope from the solar nebula gas. They show
that for a short time, a massive planet can be very bright.
Unfortunately, no evolutionary tracks giving $T_{eff}$ are
shown. Evolutionary tracks for GQ Lup and its companion calculated by
Wuchterl were presented in (Neuh\"auer et al. \cite{neuhaeuser05}) and
at this conference (see Wuchterl these proceedings). These tracks give
masses between 1 and 2 $M_{Jupiter}$ for the companion of GQ Lup.

\section{The future}

As mentioned above, the big problem is that there is not a single object
with an age of about one Myr and such a late spectral type where the
mass has been determined directly. For deriving the mass by measuring
the log(g) and $T_{eff}$, spectra with a resolution of $\Delta \lambda /
\lambda \geq 30000 $ are required. Because of the problems of the TiO
lines in the optical, such an experiment is better be carried out at
infrared wavelength. The CO-lines in the K-band could be used for
instance. However, because of the additional complication that there
could be veiling, it is required to observe not only these lines but a
much larger number of lines. While CRIRES will give the required
spectral resolution, 9 settings are required to cover the J-band, 7
settings for the H-band, and also 7 settings for the K-band. Getting the
required data with CRIRES would be time-consuming, to say the least.
Such a project thus is only feasible if an instrument like NAHUAL (see
Mart\'\i n et al, this conference) is used.

The other possibility is to determine masses directly in a binary
system. While this has not been achieved yet, it is certainly the way
to go.

% \acknowledgements


\begin{thebibliography}{}

\bibitem[2002]{barrado02}
         Barrado y Navascu\'es, D., Zapatero
         Osorio, M. R., Mart\'\i n, E.L.,
         B\'ejar, V.J.S., Rebolo, R., Mundt, R.:
         2002, A\&A~393, L85

\bibitem[2002]{baraffe02}
         Baraffe, I.; Chabrier, G.; Allard, F.; Hauschildt, P. H.:
         2002, A\&A~382, 563

\bibitem[2000]{basri00}        
         Basri, G., Mohanty, S., Allard, F., Hauschildt,
         P.H., Delfosse, X., Mart\'\i n,
         E.L., Forveille, Th., Goldman, B.: 2000, \apj 538, 36

\bibitem[2001]{batalha01}
          Batalha, C., Lopes, D.F., Batalha, N.M., 2001: \apj~548, 377
     
\bibitem[2002]{benedict02}        
         Benedict, G.F., McArthur, B.E., Forveille, T., Delfosse, X.,
         Nelan, E., Butler, R. P., Spiesman, W., Marcy, G., Goldman, B.,
         Perrier, C., Jefferys, W.H., Mayor, M., 2002: \apj~581, L115

\bibitem[1997]{burrows97}
          Burrows, A., Marley, M., Hubbard, W. B., Lunine, J. I.,
          Guillot, T., Saumon, D., Freedman, R., Sudarsky, D., Sharp, C.:
          1997, \apj~491, 856

\bibitem[2000]{charbonneau00}
         Charbonneau, D., Brown, T.M., Latham, D.W., Mayor, M.:
         2000, \apj~529, L45

% BD planet
\bibitem[2005a]{chauvin05a}         
         Chauvin, G., Lagrange, A.-M., Dumas, C., Zuckerman, B.,
         Mouillet, D., Song, I., Beuzit, J.-L., Lawrance, P. 2005: 
         A\&A~438, 25
 
% AB Pic
\bibitem[2005b]{chauvin05b}         
         Chauvin, G., Lagrange, A.-M., Zuckerman, B., Dumas, C.,
         Mouillet, D., Song, I., Beuzit, J.-L., Lawrance, P., Bessel,
         M. 2005: A\&A~438, L29

\bibitem[2002]{franco02}
         Franco, G. A. P., 2002: MNRAS~331, 474

\bibitem[2005]{guenther05}
          Guenther, E.W., Paulson, D.B., Cochran, W.D., Patience, J.,
          Hatzes, A.P., Macintosh, B. 2005: A\&A~442, 1031

\bibitem[2004]{golimowski04}
        Golimowski, D.A., Leggett, S.K., Marley, M.S., Fan, X., Geballe, T.R., 
        Knapp, G.R. et al.: 2004, AJ~127, 3516

\bibitem[2004]{hubickyj04} 
         Hubickyj, O., Bodenheimer, P., \& Lissauer, J.~J.: 2004, 
         Revista Mexicana de Astronomia y Astrofisica Conference Series 22, 83

\bibitem[1993]{hughes93}
         Hughes, J., Hartigan, P., Clampitt, L.: 1993, AJ~105, 571

\bibitem[1999]{kirkpatrick99}           
         Kirkpatrick, J.., Reid, I.N., Liebert, J., Cutri,
         R.M., Nelson, B., Beichman, Ch.A., Dahn, C.C.,
         Monet, D.G., Gizis, J.E., Skrutskie, M.F.  1999: \apj~519, 802

\bibitem[2000]{kirkpatrick00}         
         Kirkpatrick, J.D., Reid, I. N., Liebert, J., Gizis,
         J.E., Burgasser, A.J., Monet, D.G., Dahn, C.C.,
         Nelson, B., Williams, R.J. 2000: AJ~120, 447

\bibitem[1998]{knude98}
         Knude, J., H\o g, E. 1998: A\&A~338, 897

\bibitem[2003]{lineweaver03}
         Lineweaver, Ch. Grether, D. 2003:  \apj~598, 1350L

\bibitem[2003]{liu03}
         Liu, M.C., Najita, J., Tokunaga, A.T.:  2003, \apj 585, 372

\bibitem[2003]{luhman04}
        Luhman, K.L. 2004, \apj 617, 1216

\bibitem[2003]{marcy03} 
         Marcy, G., Butler, R. P., Fischer, D. A., \&
         Vogt, S. S. 2003, in ASP Conf. Ser., Scientific Frontiers in
         Research on Extrasolar Planets, ed. D. Deming, \& S. Seager (San
         Francisco: ASP)

\bibitem[1995]{mayor95}
         Mayor, M., Queloz, D.: 1995, Nature 378, 355 

% flares disks in BDs
\bibitem[2004a]{mohanty04a}
         Mohanty, S., Jayawardhana, R., Natta, A., Fujiyoshi, T., Tamura, M., Barrado y 
         Navascu\'es, D. 2004a: \apj 609, L33

\bibitem[2004b]{mohanty04b}
        Mohanty, S., Basri, G., Jayawardhana, R., Allard, F., Hauschildt, P., Ardila, D.
        2004b: \apj 609, 854

\bibitem[2004c]{mohanty04c}
         Mohanty, S., Jayawardhana, R., Basri, G.: 2004c, \apj 609, 885
          
\bibitem[2005a]{mohanty05a}
         Mohanty, S., Jayawardhana, R., Basri, G.: 2005, \apj 626, 498

\bibitem[2005b]{mohanty05b}
         Mohanty, S. these proceedings

\bibitem[2005]{muzerolle05}
         Muzerolle, J., Luhman, K.L., Brice\~no, C., Hartmann, L.,
         Calvet, N.: 2005, \apj 625, 906

\bibitem[2003]{mclean03}
         McLean, I.S., McGovern, M.R., Burgasser, A.J.,
         Kirkpatrick, J.D., Prato, L., Kim, S.S.
         2003: \apj 596, 561

\bibitem[2005]{mugrauer05}
        Mugrauer, M., Neuh\"auser, R., AN submitted

\bibitem[2000]{nakajima00}
          Nakajima, Y., Tamura, M., Oasa, Y., Nakajima, T.: 2000, AJ 119, 873

\bibitem[2003]{neuhaeuser03}
         Neuh\"auser, R., Guenther, E.W., Alves, J., Hu\'elamo, N., Ott, Th., 
         Eckart, A. 2003: AN~324, 535

\bibitem[2005]{neuhaeuser05}
         Neuh\"auser, R., Guenther, E.W., Wuchterl, G., Mugrauer, M., Bedalov, A., 
         Hauschildt, P.H., 2005: A\&A 435, L13

\bibitem[2004]{natta04}
         Natta, A., Testi, L., Muzerolle, J., Randich, S., Comer\'on, F., Persi, P.:
         2004, A\&A, 424, 603

\bibitem[2001]{reid01}           
         Reid, I. N., Burgasser, A.J., Cruz, K.L., Kirkpatrick,
         J. D., Gizis, J.E.: 2001, AJ 121, 1710

\bibitem[2003]{rice03}
         Rice, W.K.M., Armitage, P.J., Bonnell, I.A., Bate, M.R., Jeffers, S.V., Vine, S.G.:
         2003, MNRAS 346, L36

\bibitem[2003]{satori03}
        Sartori, M.J., Lepine, J.R.D., Dias, W.S.: 2003, A\&A 404, 913

\bibitem[2000]{teixeira00}
        Teixeira, R., Ducourant, C., Sartori, M. J., Camargo, J. I. B.,
        P\'eri\'e, J.P., L\'epine, J.R.D., Benevides-Soares, P.,
        2000: A\&A 361, 1143 

\bibitem[1999]{zeeuw99}
         de Zeeuw, P.T., Hoogerwerf, R., de Bruijne, J.H.J., Brown, A.G.A.,
         Blaauw, A.: 1999, AJ 117, 354

\bibitem[2001]{zucker01}
         Zucker, Sh., Mazeh, T.: 2001, \apj 562, 1038

\end{thebibliography}
\end{document}